\documentclass[preprint,showkeys,preprintnumbers,amsmath,amssymb]{revtex4}

\usepackage{graphicx}% Include figure files
\usepackage{dcolumn}% Align table columns on decimal point
\usepackage{bm}% bold math
\usepackage{epsfig}

\def\n{{\bf{\hat{n}}}}
\newcommand{\be}{\begin{equation}}
\newcommand{\e}{\end{equation}}
\newcommand{\bear}{\begin{eqnarray}}
\newcommand{\ear}{\end{eqnarray}}

\newcommand{\de}{{\rm d}}

\newcommand{\etar}{\eta_{\rm LSS}}

\def\apj{ApJ}

\def\mnras{MNRAS}
\def\prd{PRD}
 
\def\mnras{MNRAS}
\def\HI{{\rm HI}}
\begin{document}

\title{The CMBR ISW and HI 21-cm Cross-correlation Angular Power Spectrum}

\author{Tapomoy Guha Sarkar$^1$}\email{tapomoy@cts.iitkgp.ernet.in}
\author{Kanan  K. Datta$^{1,2}$}\email{kanan@cts.iitkgp.ernet.in}
\author{Somnath  Bharadwaj$^{1,2}$}\email{somnath@phy.iitkgp.ernet.in} 

\affiliation{${}^1$ Centre for  Theoretical Studies,  
I.I.T. Kharagpur, 721302, India}
  
\affiliation{${}^2$Department of Physics and Meteorology 
I.I.T. Kharagpur, 721302, India}

\begin{abstract}
The late-time growth of large scale structures  is imprinted in
 the  CMBR  anisotropy  through the Integrated Sachs Wolfe (ISW)
 effect. This is  perceived to be a very  important observational 
 probe of  dark energy. Future observations of  redshifted 21-cm
 radiation from the cosmological neutral hydrogen (HI) distribution 
 hold the potential of probing   the large scale structure over a large 
redshift  range.  
We have investigated the possibility of detecting the ISW through 
cross-correlations  between the CMBR anisotropies and redshifted 21-cm
 observations. Assuming that the HI traces the dark matter, we find
 that the ISW-HI cross-correlation angular power spectrum at an angular
 multipole $\ell$ is  proportional to the  dark matter power spectrum
 evaluated at the comoving wave number $\ell/r$, where $r$ is the
 comoving distance to the redshift from which the HI signal
 originated. The amplitude of the cross-correlation signal depends on
 parameters related to the HI distribution and the growth of
 cosmological perturbations. However, the cross-correlation  is
 extremely weak as compared to the  CMBR anisotropies and the predicted HI signal.  Even in an ideal situation, the  cross-correlation signal is smaller than the  cosmic variance and a statistically significant detection is not very likely.
 \end{abstract}

\maketitle

\section{Introduction}

In recent times, a host of independent observations, like Supernova-Ia
\cite{riess,perlmutter},  galaxy surveys \cite{tegmark} and Cosmic
Microwave Background (CMB) anisotropies \cite{dunkley,komatsu},  have
indicated
that the expansion of the Universe is accelerating \cite{Peebles}. This can be
explained by a dark energy component, with an equation of state
$p/\rho = w ( < - \frac{1}{3})$.   The cosmological constant,  $
\Lambda$, has emerged as a strong candidate for dark energy, as
various observations \cite{komatsu} constrain $w$ to be be close to $-1$.

An indirect effect of $\Lambda$ is that it causes a decay of the
gravitational potential, when the universe evolves from the matter
dominated to the  the  dark energy dominated era. This 
 generates a weak anisotropy in the CMB temperature fluctuation,
 through the Integrated Sachs Wolfe (henceforth ISW) effect  \cite{sachsW}.  
A non-flat spatial geometry would contribute to the
ISW in the same way. However, CMB data largely constrain our universe to be spatially
flat (see ref. \cite{komatsu}) so,  such effect of spatial curvature can be ignored in the
first approximation.
 The late-time evolution of the gravitational potential is sensitive to the specific dark
 energy model. Therefore, the associated ISW anisotropy, can in
 principle be used  to probe the nature of dark energy. 

It is difficult to separate the ISW signal from the primary CMB
anisotropy, because it is intrinsically weak and it appears at large
scales, where the error due to cosmic variance is large.
 
Techniques to measure the ISW, use the cross-correlation of the CMB
fluctuations, with fluctuations of some tracer of the large scale
structure at a later redshift. Fluctuations in the primary CMB field
and in the tracer are uncorrelated, so that this method allows one to
single out  the contribution solely due to the ISW. It is
also important to note here, that the foregrounds and noise are not
correlated between independent 
random fields. Recently  ISW-large scale structure  and ISW-weak lensing cross-correlations have been
studied extensively (see ref.\cite{Scranton,Afshordi,Boughn,Giannantonio,Padmanabhan,torres,Ho,Gian2}).
These studies look at a median $z \sim  1.5 $ and are in agreement
with the $\Lambda$CDM  model at $\ 2\sigma \sim 3\sigma$
levels.

  Observations of  redshifted $21 \,{\rm cm}$ radiation of the 
spin-flip hyperfine transition from neutral hydrogen
  (HI) have the potential of 
probing the universe over a wide range of redshifts ($200 \ge z \ge 0$):
from the dark ages to to the present epoch (eg.  \cite{bali,furla,antony}).
Recently, radio-optical cross-correlation study has  detected a
  positive correlation between the optical galaxies (6dFGS) and
HI fluctuations \cite{Pen et al}. 
This vindicates the theoretical predictions \cite{bns,bs} about the
possibility of 
  using  HI distribution statistically, as a  probe of the large scale
  structure, without the need to resolve individual galaxies.  
 Cross-correlation technique using the HI 21-cm radiation as one of the 
fields has been considered for the study of cosmic 
reionization \cite{adshead,Alva,Gian3,slosar,lidz}.

In this paper we study the use of diffused  HI as a tracer of the
large scale structure to probe dark energy induced ISW effect. 
We look at the cross correlation between the post-reionization ( $ z
\lesssim  6 $) fluctuations in the HI brightness temperature and the
CMB. 
                                
Redshifted   $21 \, {\rm cm}$ observations  of neutral HI allow us
to probe the universe as a function of redshift. The advantage of using
HI tomography is that,  we can probe the late-time cosmic history
continuously over a range of redshifts.  
Radio telescopes (eg. currently functioning GMRT
\footnote{http://www.gmrt.ncra.tifr.res.in/} and upcoming  MWA 
\footnote{http://www.haystack.mit.edu/ast/arrays/mwa/} \& LOFAR
\footnote{http://www.lofar.org/}) 
 are aimed to map the
large-scale distribution of HI  at high redshifts. 
At redshifts $0 \le z \le 3.5$  we have 
$\Omega_{\rm gas} \sim 10^{-3}$ (for details see
 \cite{peroux,lombardi,lanzetta}). This implies that the mean neutral
fraction of the hydrogen gas is $ \bar{x}_{\HI}=50\,\,\Omega_{\rm gas}
h^2 (0.02/\Omega_b h^2) =2.45 \times 10^{-2}$, which we assume is a
 constant over the the entire redshift range $z \le 6$.

The redshifted $21 \, {\rm cm}$ radiation seen in emission in this redshift 
range, from individual clouds is rather weak ($ < 10 \,\mu{\rm Jy}$). This
 makes its detectability dubious, with existing observational
 facilities. (There might be considerable  magnification caused by
 gravitational lensing \cite{saini} which may enhance detection chances).
Statistical distribution of HI however produces a weak 
background in radio observations. This radiation has the information
about the HI fluctuations in probed redshift range \cite{bns,bs}.
CMB map of a large portion of the 
sky and a corresponding HI map would allow us to compute the
cross-correlation power spectrum and hence independently quantify the
cosmic history at redshifts $z \le 6$.

\section{Formulation}

The  CMB brightness temperature fluctuation along the  direction of
the unit vector $ \n$ is described by 
 \cite{Hur,Subramanian}
\be
\Delta T({\bf{\hat{n}}})  = T \left\{
\left( \frac{1}{4} \delta_R + {\bf v \cdot n} + \Phi\right)_{\rm LSS}
+ \int_{\eta_{\rm LSS}}^{\eta_0} \de \eta \, [\dot{\Phi} +
  \dot{\Psi}] \right\} \,.
\e
where $T$ is the CMB temperature at present. Here, under the assumption of 
instantaneous recombination, the Sachs 
 Wolfe effect (first  term)  is evaluated at the last scattering
 surface (LSS) and the ISW effect (second term) is
 integrated from  the LSS to the present epoch. 
 The scalar potentials  $ \Phi$ and $\Psi$ are the metric
 perturbations in the conformal Newtonian gauge \cite{Bert,Br}, 
 the dots refer to differentiation with respect to the conformal time
 $\eta $ and we shall use 
 $ r = \eta_0 -\eta $ to denote the comoving distance to the 
 conformal time $\eta$. 

In the absence of anisotropic stress we have $ \Phi = \Psi$
\cite{Bert} and  the ISW term is 
\be
 \Delta T(\n)^{\rm ISW} = 2 T \int_{\etar}^{\eta_0} \de
 \eta   \dot{\Phi}(r \n,\eta). 
\e
Expanding this in  the basis of spherical
harmonics 
\be
\Delta T({\bf{{\n}}})^{\rm ISW} = \sum_{\ell,m}^\infty a_{\ell m}^{\rm
  ISW} Y_{\ell m}({\n})  
\e
and using the identity 
\be
 e^{i{\mathbf{k}}\cdot {\mathbf{n}} r}  = 4\pi \sum_{\ell,m}{
   (-i)}^{\ell} j_{\ell} (kr)Y_{\ell m}^*({\bf{\hat{k}}}) Y_{\ell m}({\bf{{\n}}})
\e
we have

\be
a_{\ell m}^{\rm ISW} = 8\pi T {(-i)}^{\ell} \int
\frac{d^3{\mathbf{k}}}{{(2\pi)}^3} 
 \int_{\etar}^{\eta_0} \de \eta \,
\dot{\tilde{\Phi}}({\mathbf{k}}, \eta) j_{\ell}(kr)Y_{\ell m}^*({\bf{\hat{k}}}) 
\label{eq:a3}
\e
where $ {\tilde{\Phi}}({\mathbf{k}}, \eta) $ 
is the Fourier transform of $ \Phi({\mathbf{r}}, \eta) $, and $j_\ell(x)$
is the spherical Bessel function.

For sufficiently sub-horizon scales the gravitational potential can be
related to the matter density fluctuations $ \delta$ via the Poisson 
equation. In 
Fourier space this takes the form 
\be
 {\tilde{\Phi}}({\mathbf{k}},\eta )= - \frac{3}{2} \frac{H_{0}^2}{c^2}
 \frac{\Omega_{m0}}{k^2} \frac{\delta({\mathbf{k}},a)}{a} 
\e

Further, retaining only the growing mode of density perturbations 
 $ \delta({\mathbf{k}},a) = \delta({\mathbf{k}}) D_{+}(a)$  we have 
\be
\dot{\tilde{\Phi}}({\mathbf{k}}, \eta)=\frac{(f-1) \dot{a} }{a}
\tilde{\Phi}({\mathbf{k}}, \eta)
\e
where 
\be
 f = \frac{d ln D_{+}}{d(ln a)}\,
\e
which we use in eq. (\ref{eq:a3}) to calculate $a^{ISW}_{\ell m}$.

The HI 21-cm brightness temperature fluctuations from redshift $z_{HI}$
can, in Fourier space,   be written as \cite{bharad04}
\be
\Delta_{HI} ({\mathbf{k}})
=\bar{T}\bar{x}_{HI} ( b + f\mu^2)\delta({\mathbf{k}},a)
\label{eq:a3.1}
\e
where  $\bar{x}_{HI}$  is
the mean HI  fraction, $\mu
={\bf{\hat{k}}}\cdot{\bf{\hat{n}}}$ 
and 
\be
\bar{T}(z)=4.0 \, {\rm mK}\,\,(1+z)^2  \, \left(\frac{\Omega_{b0}
  h^2}{0.02}\right)  \left(\frac{0.7}{h} \right) \frac{H_0}{H(z)}
\e
Here it has been assumed that the HI traces the underlying dark matter 
distribution with a possible bias $b$. On the large scales under 
consideration, where the matter fluctuations
are in the linear regime, it is reasonable to assume that the baryonic
matter follows the underlying dark matter
distribution. The term $ f \mu^2$ has its origin in the  HI peculiar
velocities \cite{bns,bharad04} which have also been assumed to be
caused by the dark matter fluctuations. It should be noted that all
the terms on the {\it rhs.} of equation  (\ref{eq:a3.1}) are to be
evaluated at the redshift $z_{HI}$ at which  the HI signal
originated. Note that one should include
a normalized window function $W(z)$  in eq. (\ref{eq:a3.1}) describing
the spectral response of an instrument \cite{slosar}. On scales of our interest ($\ell
\lesssim 100$), the spectral resolution of the instrument can however be
assumed to be much smaller than the features in the HI signal
\cite{zaldafur} and  W(z) can be approximated by a Dirac delta
function, so that  eq. (\ref{eq:a3.1}) is, a reasonably good approximation.

Expanding the HI signal in terms of spherical harmonics and proceeding
as before we get 
\be
a_{\ell m}^{\rm HI}   {=}  4\pi \bar{T}(z)\bar{x}_{HI} {(-i)}^{\ell}
\int
\frac{d^3{\mathbf{k}}}{{(2\pi)}^3} 
\delta({\mathbf{k}},a)\mathcal{I}_{\ell}(kr)Y_{\ell m}^*({\bf{\hat{k}}}) \,
\label{eq:a4}
\e
where
\be
\mathcal{I}_{\ell}(x) {=}  b j_{\ell}(x)- f \frac{d^2 j_\ell}{dx^2} \,.
\e

We use equations (\ref{eq:a3}) and  (\ref{eq:a4}) to calculate
$\mathcal{C}^{HI-ISW}_{\ell}$  the
cross correlation angular power spectrum between the 
HI 21 cm brightness temperature signal and the CMBR ISW signal defined
through 
\be
\langle  a_{\ell m}^{\rm ISW} a_{\ell'm'}^{*\rm HI} \rangle  =
\mathcal{C}^{HI-ISW}_{\ell} 
\delta_{\ell \ell'}\delta_{mm'} 
\e
Note that $\mathcal{C}^{HI-ISW}_{\ell}$ also depends on $z_{HI}$ the
redshift from which the HI signal originates, or equivalently on
$\nu=1420 \, {\rm MHz}/(1+z_{HI})$ the frequency of  the HI
observations, but we do not show this explicitly here. We obtain 
\be
\mathcal{C}_{\ell}^{HI-ISW} {=} \mathcal{A}(z_{HI}) \int dk \left [ P(k)
  \mathcal{I}_\ell(k r_{HI}) \int_{\etar}^{\eta_0} \de \eta F(\eta)
  j_{\ell}(kr)\right]
\label{eq:cl1}
\e 
where $P(k)$ is the present day dark matter power spectrum, 
\be
 \mathcal{A}(z) = -\bar{T}(z)\bar{x}_{HI}
 D_{+}(z)  \frac{6 H_{0}^3 \Omega_{m0}}{\pi c^3}  
\e
and 
\be
F(\eta) =  \frac{D_{+} (f-1) H(z)}{H_0}
\e

For large $\ell$ we can use the Limber approximation \cite{limber, Afshordi}
which allows us to replace the spherical Bessel functions by a  Dirac
deltas $\delta_D(x)$ 
\be
j_{\ell}(kr) \approx  \sqrt{\frac{\pi}{2\ell+1}} \delta_D(\ell + \frac{1}{2} - kr)
\e
whereby  the angular cross-correlation power spectrum 
takes the simple form 
\be
   \mathcal{C}_{\ell}^{HI-ISW}  \approx \frac{\pi  \mathcal{A} (b+f) F } {2\ell^2}
P(\frac{\ell}{r})
\label{eq:fnl}
\e
where $P(k)$ is the present day dark matter power spectrum and all the
other terms on the {\it rhs.} are evaluated at 
$z_{HI}$.

We also have, for comparison, the  HI-HI angular power spectrum
$C_{\ell} ^{HI}(\Delta \nu) $ 
\cite{datta1}, which  describes the statistical properties of HI
fluctuations at two redshifts (corresponding frequencies being  $\nu$
and $\nu + \Delta \nu$). Using the `flat sky' approximation 
\cite{datta1} we have
$ C_l^{HI} (\Delta \nu) $ is given by
\begin{equation}
C_\ell^{HI}(\Delta \nu ) =
\frac{\bar{T}^2~ }{\pi r_{\nu}^2} \bar{x}^2_{\HI}   D_{+}^2
\int_{0}^{\infty} {\rm d} k_{\parallel} \, 
\cos (k_{\parallel}\, r'_{\nu}\, \Delta \nu) \,  \left[  b+  f \mu^2
\right]^2 P(k) \,
\label{eq:fsa} 
\end{equation}
where $ r$ is the comoving distance corresponding to the
redshift $z_{HI}$ or equivalently frequency $\nu = 1420 {\rm
  MHz}/(1+z_{\HI})$,  $r_{\nu}' = d r_{\nu}/d \nu
 $  and $k=\sqrt{k^2_{\parallel}+(l/r)^2}$. 

 The function $C_{\ell}^{HI}(\Delta\nu)$ is a direct observational
 estimator of the HI fluctuations at redshift $z_{\HI}$. This  
 does not  require  us to assume  an underlying cosmological 
 model (eg. \cite{ali08}). Here we have taken the special case where $
 \Delta\nu = 0$  and we shall henceforth refer to the corresponding
 power spectrum as $C_{\ell}^{HI}$.

\section{Results}

\begin{figure}
\begin{center}
  \mbox{\epsfig{file=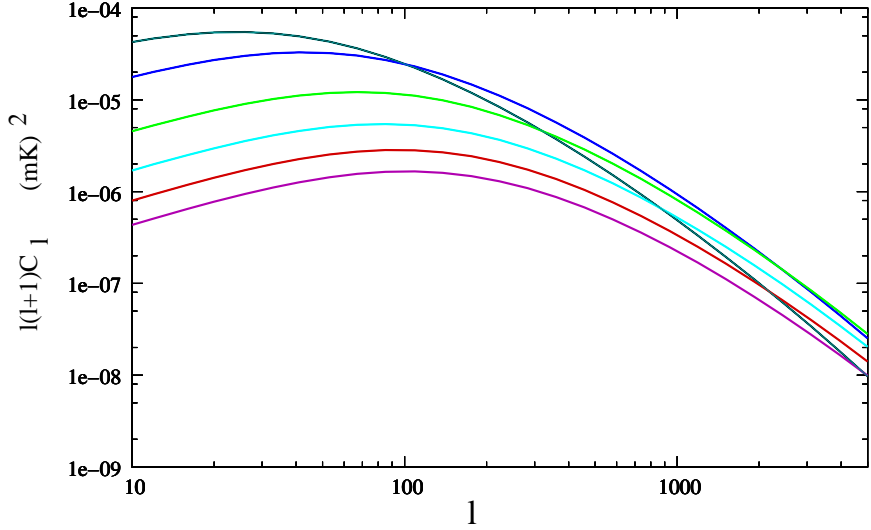,width=0.7\textwidth,angle=0}}
\caption{The HI-ISW angular power spectrum  for  redshifts $z=
  0.5. 1.0, 2.0, 3.0, 4.0, 5.0$ (top to bottom).} 
\label{fig:kappa}
\end{center}
\end{figure}

\begin{figure}
\begin{center}
\mbox{\epsfig{file=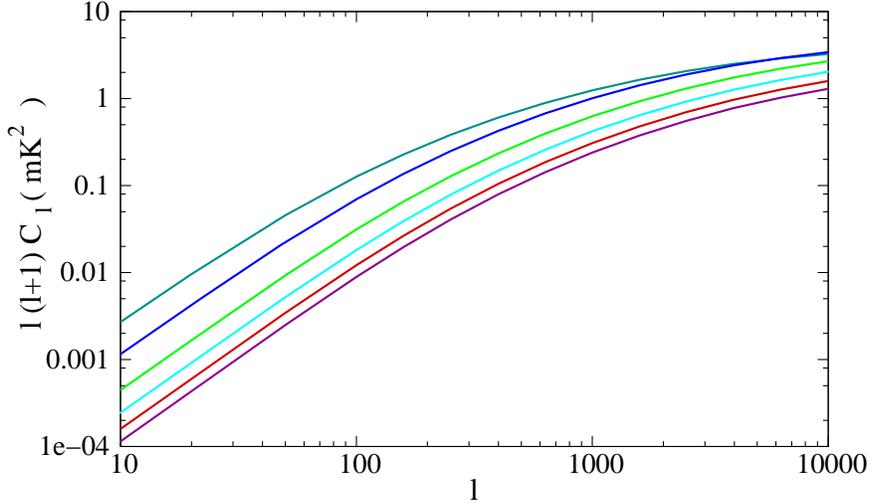,width=0.7\textwidth,angle=0}}
\caption{The HI angular power spectrum  $C_\ell^{HI}$  at redshifts
  $z=0.5. 1.0,  2.0, 3.0, 4.0,  5.0$ (top to bottom). }
\label{fig:cl}
\end{center}
\end{figure}

 Figure~\ref{fig:kappa} and Figure~\ref{fig:cl} respectively show the
 predicted  HI-ISW cross-correlation angular power spectrum
 $\mathcal{C}^{HI-ISW}_\ell$ and the HI-HI angular power spectrum
 $C_\ell^{HI}$ for a few redshifts $z_{HI}$ in the range $0.5 \le z_{HI}
 \le 5$. We have used equations \ref{eq:cl1} and \ref{eq:fsa} to
 calculate the cross-correlation angular power spectrum and HI power
 spectrum respectively. The approximated equation (\ref{eq:fnl}) is
 useful for  qualitative description of  the results. We have assumed the currently favored  $\Lambda$CDM
 cosmological model 
 with parameters
 $(\Omega_{m0},\Omega_{\Lambda0},h,\sigma_8,n_s)=(0.28, 0.72, 0.7,
 0.82, 0.97)$ \cite{dunkley,komatsu}. The bias, $b$ for the post
 reionization HI on large scales is assumed to be
 linear. We have taken $ b = 1 $ as the fiducial model. However, it is
 important to note that HI in the post reionization epoch is assumed
 to be distributed in high column density clouds which could be more
 biased with respect to the underlying cold dark matter distribution.

The shape ($\ell$ dependence)  of the  cross-correlation signal reflects
 the shape of the matter power spectrum $P(k)$ (eq. \ref{eq:fnl}). 
We find a peak in $\mathcal{C}^{HI-ISW}_\ell$  at $\ell=r k_{eq}$, where $k_{eq}$ is the wave vector corresponding to the 
 matter radiation equality. For different redshifts $z_{HI}$ 
the $\ell$ value corresponding to
 this peak scales as $\ell \propto r$, the comoving distance to the
 redshift $z_{HI}$.  

The amplitude of the cross-correlation signal $\mathcal{C}^{HI-ISW}_\ell$ 
depends on a product of various terms some of which
($\bar{T},\bar{x}_{HI}, b$)  depend on the HI distribution and
others $(D_+,f,H)$ which depend on the cosmological model. 
The dimensionless term $f$ quantifies the growth of the dark matter 
perturbations, and
the ISW effect is proportional to $f-1$. We have $f=1$ in
cosmological models with no dark energy, and we do not expect to have
any ISW effect in such models.  The term $f-1$ is a sensitive
probe of  dark energy. 
The amplitude of  $\mathcal{C}^{HI-ISW}_\ell$
contains this information combined with unknown parameters related to
the HI distribution.  It has been recently proposed that observations
of the HI fluctuations at low $z$ can be used to estimate cosmological
parameters \cite{BSS,wyithe}. It is in principle possible to combine
observations of $\mathcal{C}^{HI-ISW}_\ell$ and
$\mathcal{C}^{HI}_\ell$ to jointly estimate parameters of the HI
distribution and the background cosmological model.  

\section{Detectability and Conclusions}

Here we estimate  the viability of detecting the HI-ISW
cross-correlation signal. 
The cosmological HI signal is weak and buried under the foregrounds 
which are orders of magnitude higher than
the signal \cite{bali,santos2,maq,datta1,ali08}. This is a 
serious observational problem for
auto-correlation studies involving the $21$cm radiation.
One may separate the foreground components by noting that HI 
signal (a line emission)  decorrelates beyond a certain frequency
separation whereas the foregrounds remain correlated over large
frequency separations. We shall subsequently assume that foregrounds
have been removed. Moreover, the cross-correlation signal is less affected 
by foregrounds and other systematics. This is because, many of the
foregrounds and noise are  expected to be uncorrelated between the two maps.

The uncertainty in estimating the cross correlation signal 
is the sum, in quadrature, of the instrumental noise and the cosmic
variance.  While the system noise can, in principle, be reduced by
increasing the duration of the observation the cosmic variance sets a
fundamental limit in deciding whether the signal can  
at all be detected or not. 

\begin{figure}
\begin{center}
  \mbox{\epsfig{file=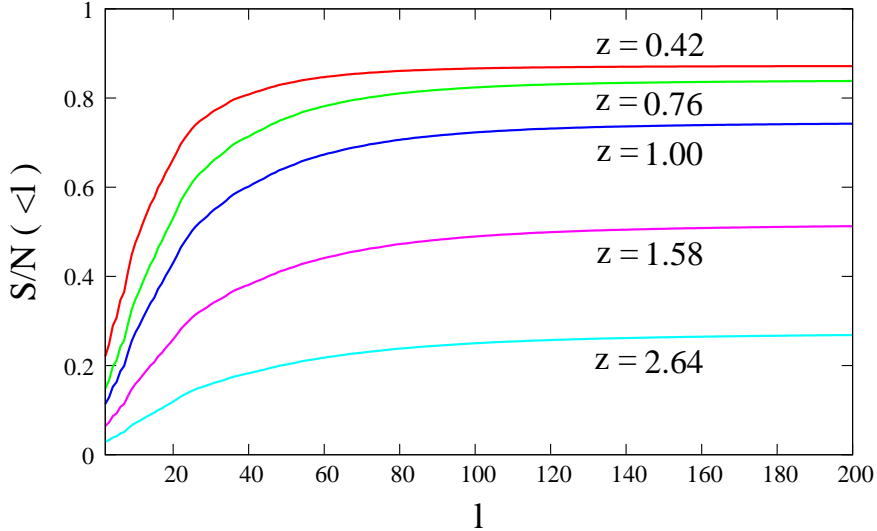,width=0.7\textwidth,angle=0}}
\caption{The cumulative ${\rm S/N}$  by collapsing all multiploes less
  than $\ell$ for  different redshifts.}
\label{fig:signoise}
\end{center}
\end{figure}

The cosmic variance of the cross-correlation angular power spectrum 
$\mathcal{C}^{HI-ISW}_\ell$ is \cite{Afshordi}
\be 
\sigma^2=\frac{C^{CMB}_{\ell} C^{HI}_\ell} {(2\ell + 1 )\sqrt{N_c} f_s
  \, \Delta \ell} 
\e
 where $C^{CMB}_{\ell}$ is the CMB angular spectrum for which we have
 used the WMAP5  results \footnote{http://lambda.gsfc.nasa.gov/}, $\Delta \ell$ is the width of bands in
 $\ell$  and  $ f_s$ is the fraction of the sky 
 common to both the CMBR and HI observations. We have used $\Delta
 \ell=10$ for $\ell \le 100$ and  $\Delta \ell=100$ for $\ell > 100$,  
and have  assumed the most  optimistic possibility $f_s=1$ for 
 our estimates at redshift $z_{\rm HI}=0.5$.  
 Different frequencies channels in the frequency band of HI
 observations provide  $N_c$ independent estimates of the HI  
 signal  which cause a reduction in the cosmic variance by a factor
 $1/\sqrt{N_c}$.  Here we have assumed that the HI observations are
 carried out across a bandwidth of $32 \, {\rm MHz}$  centered around 
$z_{\rm HI}=0.5$ and the HI
 signal is assumed to be independent  at frequency separations 
 of $\sim  1  {\rm MHz}$ \cite{bs}, which gives $N_c=32$.   
Using these to estimate the signal to noise ratio ${\rm
  S/N}=\mathcal{C}^{HI-ISW}_\ell/\sigma$  we find that ${\rm S/N} <
0.45$ for  all $z_{HI}$ and $\ell$ and a statistically significant
detection is not possible in such cases. It is possible to
increase  ${\rm S/N}$ collapsing  the signal at different multipoles
$\ell$.   To test if a statistically significant detection is thus
feasible we have collapsed all multipoles less than $\ell$ to evaluate
the cumulative ${\rm S/N}$ defined as \cite{cooray2,adshead}
\be
{\left (\frac{S}{N} \right)}^2 = \sum_{\ell} \frac{(2\ell + 1 )\sqrt{N_c} f_s
  \left(\mathcal{C}_l^{HI-ISW}\right)^2}{C^{CMB}_lC^{HI}_l}. 
\e 
Results are shown in Figure \ref{fig:signoise} for various 
redshifts ($ 0.4 < z  <3$). We find  that the contribution in the 
cumulated $S/N$ comes from
$\ell<50$ at all redshifts that we have  considered. The
cross-correlation signal is largest at ($z \sim 0.4$) and is negligible
for ($ z > 3$). We further find that although there is an increase in ${\rm
  S/N}$ on collapsing the multipoles it is still less than unity. This
implies that a statistically significant detection is still not possible. Thus, probing a thin shell of HI doesn't allow us to detect a cross correlation, the signal being limited by the cosmic variance. 

\begin{figure}
\begin{center}
  \mbox{\epsfig{file=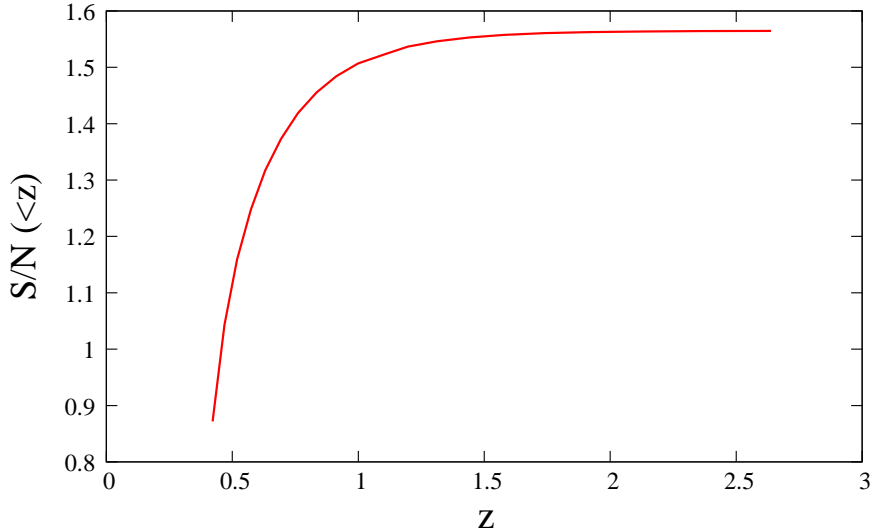,width=0.7\textwidth,angle=0}}
\caption{The cumulative ${\rm S/N}$ on combining data upto a certain redshift.
  }
\label{fig:sin}
\end{center}
\end{figure}

 $21$ cm observations have the advantage that one may probe various
redshifts by tuning the frequency of radio observations. This enables us to optimally combine the signal from a large
number of thin shells over a continuous range of redshifts. We have
considered a range of redshifts ($ 0.4 < z < 3 $ or  $ 1000 > \nu  > 350 $) and
combined the signal for independent observations at $32$ MHz
separations in  this range. The S/N
cumulated upto a certain redshift is shown in Figure \ref{fig:sin}. This 
indicates an
increase in the S/N. A cumulated S/N of $ \sim 1.6 $ is attained for redshift 
upto $z = 2$
and there is hardly any increase in S/N on cumulating beyond this redshift.
This is reasonable because the contribution from the ISW effect
becomes smaller beyond the redshift $z>2$. This S/N is the
theoretically calculated value for an ideal situation and is unattainable for
most practical purposes. Incomplete sky coverage, and foreground
removal issues would actually reduce the S/N and attaining a
statistically significant level is not  feasible.
We conclude that, within the paradigm of $\Lambda CDM$ cosmology, though there is a weak positive correlation between
the CMBR ISW and HI, the signal is much weaker than the
individual auto-correlations and a detection is quite unlikely.  Certain modified gravity models (eg.\cite{bertzukin,song}) may allow  the quantities  $(D_+,f,H)$
to be  different from what they are in the $\Lambda CDM$
model (considered here)\cite{sudaqua} and may lead to an increase of the $S/N$.
However, since  the cross-correlation signal is significant only at large scales we don't expect the $S/N$ to be much different from the $\Lambda CDM$ predictions.

\acknowledgments

We thank Jasjeet Bagla and T. Roy Choudhury for useful discussions. TGS 
would like to acknowledge
financial support from Board of Research in Nuclear Sciences (BRNS), 
Department of Atomic Energy (DAE), Government of India through the project
2007/37/11/BRNS/357. KKD would like to acknowledge
financial support from Council of Scientific and Industrial Research
(CSIR), India through senior research fellowship.

\bibliography{apssamp}

\end{document}